\documentstyle[11pt,aaspp4]{article}

\begin{document}
\title{Kinematic Effects of Tidal Interaction on Galaxy Rotation Curves
\footnote{Some observations reported in this paper were obtained
at the Multiple Mirror Telescope Observatory, a facility operated
jointly by the University of Arizona and the Smithsonian Institution.
}}
\author{Elizabeth J. Barton}
\affil{Harvard-Smithsonian Center for Astrophysics}
\authoraddr{ Mail Stop 10, 60 Garden St. Cambridge, MA 02138,
	(email: ebarton@cfa.harvard.edu)
}
\and
\author{Benjamin C. Bromley}
\affil{Department of Physics, University of Utah}
\authoraddr{201 JFB, Salt Lake City, UT 84112,
	(email: bromley@physics.utah.edu)
}
\and
\author{Margaret J. Geller}
\affil{Harvard-Smithsonian Center for Astrophysics}
\authoraddr{Mail Stop 19, 60 Garden St., Cambridge, MA 02138,
	(email: mgeller@cfa.harvard.edu)
}

\begin{abstract}

We use self-consistent N-body models, in conjunction with models
of test particles moving in galaxy potentials, to explore the
initial effects of interactions on the rotation curves of spiral
galaxies.  Using nearly self-consistent disk/bulge/halo galaxy
models (Kuijken \& Dubinski 1995), we simulate the first pass of
galaxies on nearly parabolic orbits; we vary orbit inclinations, 
galaxy halo masses and impact parameters.  For each simulation,
we mimic observed rotation curves of the model galaxies. 
Transient interaction-induced features of the curves include distinctly
rising or falling profiles at large radii and pronounced bumps in
the central regions. Remarkably similar features occur in our
statistical sample of optical emission-line rotation curves of
spiral galaxies in tight pairs and n-tuples. 

\end{abstract}

\keywords{galaxies: interactions --- galaxies: kinematics and dynamics}

\section{Introduction}

Observations of interacting pairs of galaxies in the nearby universe
reveal a rich set of phenomena, including starbursts, tidal tails,
streams, bridges and shells, which arise as
companions perturb or even disrupt
the galaxies (see Barnes \& Hernquist 1992; Sanders \& Mirabel 1996).  These
features, if unique to interacting systems, may offer probes of the
extent of dark halos in individual galaxies, the rate of galaxy
mergers, and the evolution of galactic structure.  Thus, analysis of
interacting galaxies may provide strong constraints for cosmological
models: the mass in dark halos may be the single most important
contribution to the cosmological mass density parameter $\Omega$.
The galaxy merger rate is also sensitive to the form of the
primordial power spectrum and hence the nature of dark matter.
  
Complete samples of interacting galaxies are valuable for 
understanding the range of dynamical processes which are important 
in galaxy evolution, and for the constraints they can place on the 
{\em average} halo mass fraction and merger rates.
Members of a late-stage merger may be unresolved, making
morphological selection of complete samples difficult. 
Nonetheless, merging pairs identified by morphology or infrared
emission can yield interesting results (e.g. the halo mass
constraints: Dubinski, Mihos, \& Hernquist 1996; Hibbard \& Mihos 1995; 
Springel \& White 1998; ULIGs: see Sanders \& Mirabel 1996 for a review). 
 
The early stages of interaction are easier to survey and to study
statistically because the systems are resolvable galaxy
pairs. Galaxy pair studies employ a variety of selection criteria
(e.g. Vorontsov-Vel'yaminov 1959; Arp 1966; Karachentsev 1972, 1987; 
Turner 1976a, 1976b; Chengalur, Salpeter, \& Terzian 1993, 1994, 1996) and
selection effects may be difficult to model.  
A few authors have attempted statistical studies of the rotation
curves of interacting galaxies (Chengalur {\it et al.} 1994; 
Keel 1993, 1996; M\'{a}rquez \& Moles 1996).  
 
Complete redshift surveys allow detailed control of all selection
effects of complete pair samples.  In observation times that permit
construction of large samples, rotation curves provide kinematic data
over a strip of each galaxy.  Barton {\it et al.} (1998b) measured
$\sim 140$ emission line rotation curves in pairs selected from the
CfA2 redshift survey (Geller \& Huchra 1989; Huchra, Vogeley, \&
Geller 1998; Falco {\it et al.} 1998).  The data display features
which are consistent with Keel's (1993;1996) qualitative studies of
rotation curves in a sample of paired and interacting galaxies.
 
This Letter is an investigation of the signatures of interaction in
galaxy pairs during the early stages of interaction or merging.  Our
numerical work with dissipationless N-body simulations (Barton {\it et
al.}  1998a) illustrates that rotation curves of spiral galaxies in
pairs can bear specific, predictable signatures of interaction: the
flat rotation curves typically associated with these objects in
isolation acquire distinct features such as dips near the galactic
nucleus and sharply rising or falling profiles in the outer regions of
the galactic disk. These features can reflect non-circular motion and
depend on the internal structure of the individual galaxies and the
orbital parameters of the interaction.  
 
\section{Numerical Method}

The first step toward simulating galaxy pair interactions is to
construct model galaxies with disk, bulge and halo components.  We
follow the general prescription of Kuijken \& Dubinski (1995),
correcting a sign error in their equation~2. The method generates
near-equilibrium phase-space distributions of galaxies in
isolation. For the purposes of this study we select three models which
reproduce the Galactic rotation curve out to roughly 5 disk scale
lengths. These models, labeled Milky Way (MW) A, B, and D by Kuijken
\& Dubinski, have halo-to-(disk$+$bulge) mass ratios of 4:1, 8:1 and
30:1, respectively, and radial extents of 22, 30, and 73 in units of
the characteristic disk scale length. For the Milky Way, the length
unit in the models has an approximate physical value of 4.5~kpc, while
the time and velocity units discussed below correspond to 20~Myr and
220~km/s, respectively.
 
We generate realizations of the Kuijken--Dubinski MW models with
discrete particles and evolve them, both in isolation and in pairs,
using a gravity treecode (Barnes \& Hut 1986; Hernquist 1987). In the
simulations, individual galaxies consist of 50,000 or 100,000
particles; the treecode's particle-cell force calculations are
performed with opening angle $\theta \leq 0.7$, monopole cell-mass
approximations, and a Plummer smoothing parameter of $\epsilon =
0.05$. The integrator is a leapfrog algorithm with a timestep of
$\Delta t = 0.01$ units.
 
To improve our ability to trace the behavior of the outer disk
accurately, we also consider galaxy pair models which consist 
of the analytical Kuijken--Dubinski potentials moving along 
trajectories determined from
the self-consistent $N$-body simulations.  We track the kinematical
evolution of the disk components with test particles (see Dubinski,
Hernquist, \& Mihos 1997). Thus, at little computational expense we
can model well-resolved disks with $N = 100,000$ particles and 
without the effects of discreteness such as artifical disk heating by massive
halo particles.
 
A comparison between the self-consistent $N$-body simulations and the
kinematical models demonstrates that the bulk flows of disk material,
and hence rotation curves, are qualitatively similar. Even detailed,
quantitative differences are negligible unless the halo mass is
relatively high (as in the MW~D models) or unless the impact parameter
of the mass centers is less than 2~units (as measured from the initial
parabolic orbits). Thus we justify our ample use here of the
kinematical models and their more highly resolved rotation curves.

We start the galaxies on initially
parabolic orbits ($E_0=0$) which eventually lead to mergers due
to dynamical friction.  The galaxies start with dark matter halos touching
(except the large Milky Way D model),
following Dubinski {\it et al.} (1997).  We observe no significant
disturbance in the disk and bulge particles in the early timesteps,
until about $T \gtrsim 30$, indicating that results are insensitive to the
proximity of the galaxies at startup.

\section{Results}

Figs.~\ref{fig:allvary} illustrates some of 
the most dramatic kinematic effects shortly after the first pass.  
The figure shows positions and velocities of particles from the 
kinematic simulations with various inclinations, 
model halo sizes and impact parameters. We plot the galaxies and 
rotation curves at $\sim 5$--$8$
time units after closest approach, when the kinematic response is
dramatic.  We rotate some galaxies in their disks' planes to
compare responses. The figure caption gives the starting galaxy parameters
and Toomre disk orientation parameters $i$ and $\omega$ (Toomre \& Toomre
1972).  For $\omega$, we quote the argument of periapse, using bulge centers
for the galaxy positions; b is the bulge separation at closest approach.  

The galactic disks, and the orbits of the fully prograde and 
retrograde galaxies, 
are in the x-y plane, the plane of the paper in
the upper half of each square. In the lower half of each square,
we show the velocities as if we were observing the galaxy inclined only
5$^{\circ}$ from edge-on, at a distance of 77 Mpc through a 1.5 arcsecond
slit.  Thus, x-positions of the particles are plotted vs. (nearly)
y-velocities.  We incline the galaxy by this small amount in order to
mimic longslit observations of velocities near the the
projected disk major axis. 
In Fig.~\ref{fig:allvary}, the solid line is the rotation 
curve we derive from the simulation
to compare with measured optical rotation curves of galaxies. To compute
this curve, we smear each particle in the x-direction 
by a ``seeing function'', a
Gaussian of 1 arcsecond (assuming a distance of 77 Mpc). We then find the
``flux''-weighted velocity mean at various positions along the slit.  

During the simulations, non-circular motions result 
as the outer disk is tidally ripped into tails.  As the orbital pass 
progresses, the outer portions of
the curve, initially flat, are stretched into rising,
falling, or bent lines.  The rotation curve for the same galaxy at the 
same time can rise or fall, depending on the viewing angle.  
The inner disk distorts into a bar-like, elongated structure 
which has almost no effect on the center of the rotation curve, 
the solid-body portion.  For a period of time after the first pass
of strong enough interactions, the inner and outer 
portions of the galaxy appear as
distinct kinematic components --- a lump on the inside and a distorted outer
region.  This kinematic pattern resembles the two-component
kinematic structure associated with
both bars and circumnuclear gas disks (Rubin, Kenney, \& Young 1997).
The pattern does not require dissipative matter to form  --- it can
form without gas infall.  At later times, the
tails become sparse as they begin to settle back onto the galaxy.

Encounters with smaller impact parameters appear to produce narrower, 
more well-defined central bumps, as
well as richer and sometimes longer tidal tails with more dramatic
kinematic features.  Not surprisingly, prograde
encounters give rise to the most dramatic responses --- the longest tidal
tails and the most kinematic tidal ripping.  Retrograde
encounters show almost no response.  
As  Dubinski {\it et al.} (1996) note, tail formation is suppressed in 
collisions of Milky Way D models. We can measure the
true differences among the models by ``observing'' the 
simulations at a large number of random orientations for comparison
with a complete observational sample.

\section{Discussion}

\begin{figure}
\centerline{\epsfysize=7in%
\epsffile{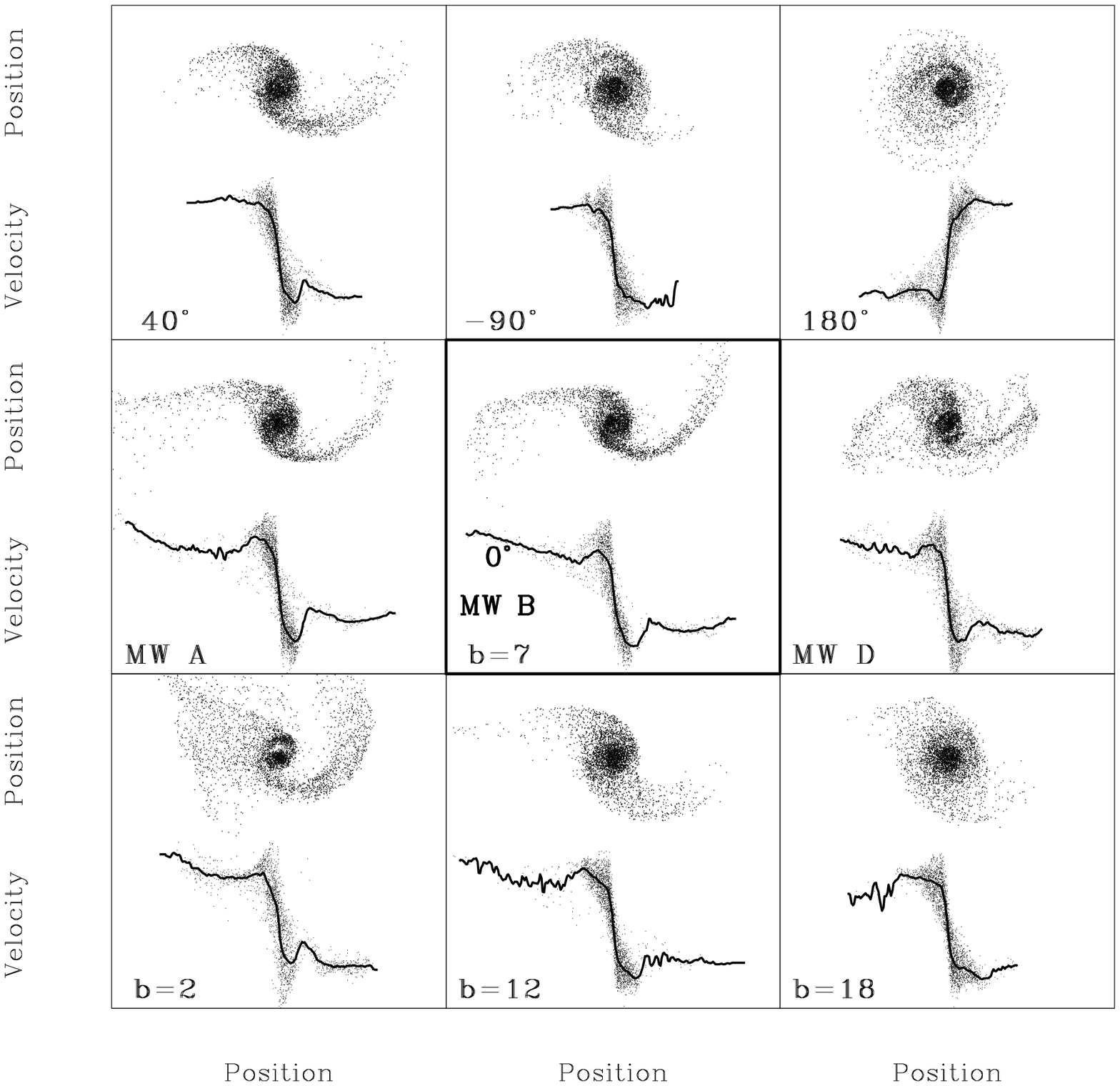}}
\caption{Galaxies and rotation curves 
from the kinematic models. In each case the simulations are
two colliding galaxies are of the same Milky Way model; we show only
one of the galaxies in each square.  The middle box is the 
fiducial simulation, a ``Milky 
Way B'' (MW B), prograde (0$^{\circ}$ inclination) galaxy with an impact
parameter of $b = 7$.  The other boxes contain variants on this
simulation.  The {\it top row} shows MW B models with impact parameters
$b=7$ and varied orbital inclinations ($i$); they
are: (a)  $i=40^{\circ}$, with Toomre disk orientation parameter 
$\omega = 30^{\circ}$ (b) $i=-90^{\circ}$, $\omega = 40^{\circ}$, and (c)
$i=180^{\circ}$ (retrograde).  The {\it middle row} shows models with
impact parameters $b=7$ and inclinations $i=0^{\circ}$ with varied 
galaxy models; they are (d) MW A (e) MW B and (f) MW D.
The {\it bottom row} shows MW B models with inclinations
$i=0^{\circ}$ and varied impact parameters; they are:
(g) b=2, (h) b=12 and (i) b=18.}
\label{fig:allvary}
\end{figure}

\begin{figure}
\centerline{\epsfysize=7in%
\epsffile{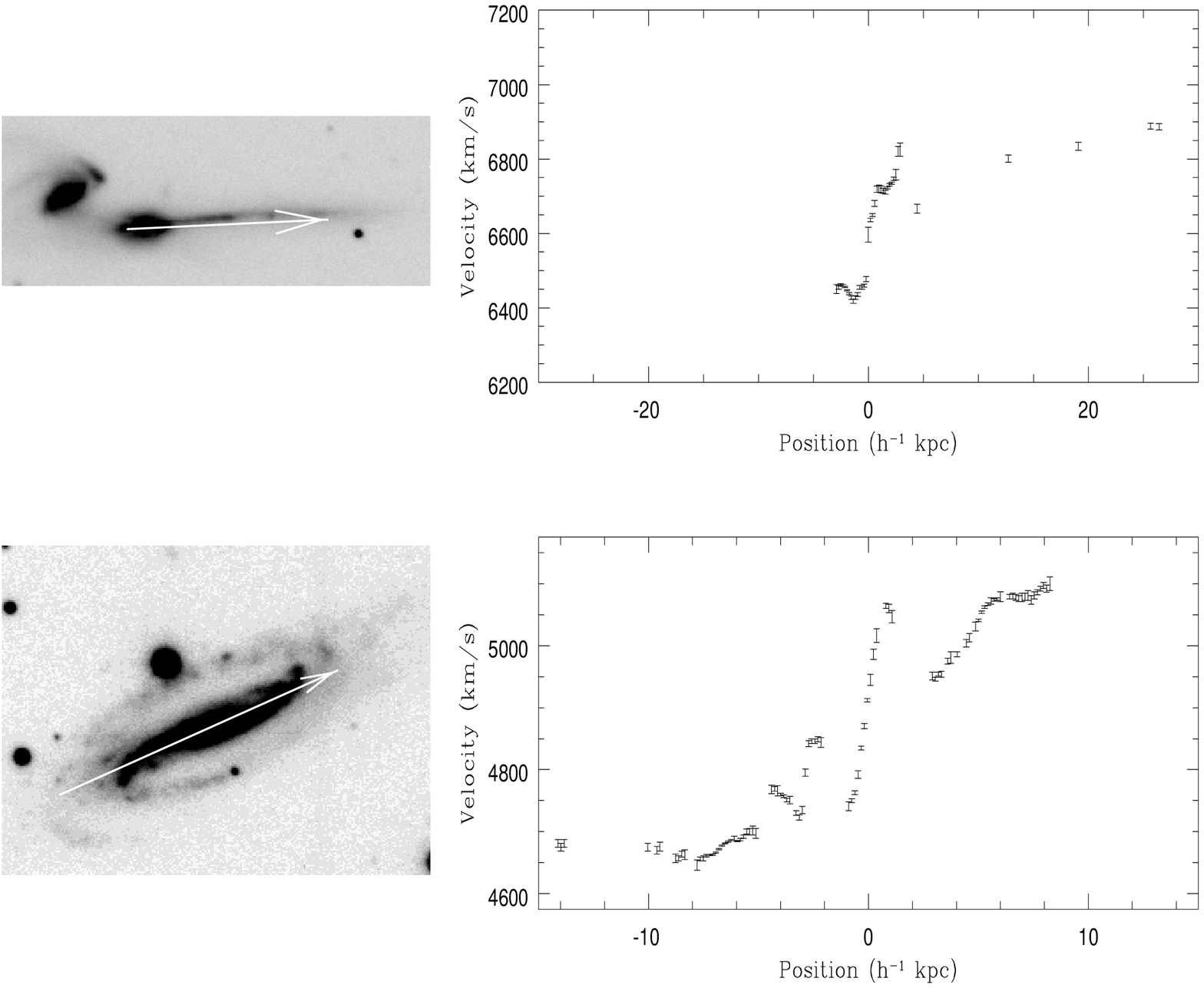}}
\caption{B Images and rotation curves 
for two galaxies in the sample:
(a) NGC 4676, ``The Mice'' and (b) UGC 484, a barred spiral.  Arrows 
show the slit position; they point in the direction of increasing
position.}
\label{fig:data1}
\end{figure}

A range of observational phenomena in our statistical sample 
of optical rotation curves of galaxies in
pairs (Barton {\it et al.} 1998b) may reflect interactions like 
those simulated in Fig.~\ref{fig:allvary}.
Fig.~\ref{fig:data1} shows striking examples of galaxies with interesting
kinematics.  
In each B image, the arrow shows the position of the spectrograph
slit; the arrow points in the direction of increasing position
on the rotation curve plot.  The curves are centered --- 
the zero position is approximately the flux center of the
continuum. 
Fig.~\ref{fig:data1}a shows NGC 4676, or ``The Mice''.  
The northern galaxy, shown to the right in our image, has one of the most 
dramatic tidal tails in our pair sample; the
rotation curve clearly shows the rising pattern.
Fig.~\ref{fig:data1}b shows UGC 484, a paired barred spiral 46 h$^{-1}$ 
projected kpc from a companion galaxy. Its rotation curve displays
the multi-component kinematic structure which may be associated with 
the bar.  The inner region shows a steep rise;  the slower
nearly solid-body rotation in the middle region extends for 
$\sim 12$ h$^{-1}$ projected kpc before the curve flattens. 

The pair sample (Barton {\it et al.} 1998b), 
consists of $\sim 140$ spiral and S0 
galaxies in spiral/spiral and spiral/elliptical pairs that are appropriate 
for longslit emission observations.  
This ``dynamical'' sample, when properly 
compared with control samples and simulations,
will enable estimation of the frequency of 
interaction-induced features.
Although we observe 
$< 5$ disk scale lengths,  we
find some curves which rise or, less frequently, fall; 
some of these curves eventually flatten. 
We also find some dips in the center which may be associated with circumnuclear
bars or gas disks (Rubin {\it et al.} 1997).  Curves of the most
distorted galaxies appear bumpy and irregular.  These phenomena
are similar to effects noted by Keel (1993, 1996).

A compelling argument linking particular features in rotation curves 
to interactions is difficult.  Statistical observational studies
can link rotation curve features to known interaction phenomena
(Keel 1993).  Simulations and models can show 
the physical basis for interaction 
features (e.g. tails, bridges, kinematic features: Toomre \& 
Toomre 1972;   Byrd 1976; Borne 1988a, 1988b; Byrd \& Klaric 1990;
Elmegreen {\it et al.} 1995a, 1995b; Combes {\it et al.} 1995; 
Dubinski {\it et al.} 1996; gas infall: 
Negroponte \& White 1983; Noguchi 1991). We extend 
this work with systematic simulations and exploration of features that might 
appear in a large sample of rotation curves of galaxies in interacting
pairs.  However, there are many aspects of galaxy interactions and
rotation curve measurement we do not model, including: (1) 
instrumental effects such as slit misalignment, (2) gas-related 
phenomena such as gas infall and star formation (e.g. Barnes \& 
Hernquist 1992; Mihos \& Hernquist 1996) (3) lumpy, non-uniform emission 
and dust which affect rotation curve measurement, (4) varied starting
galaxy potentials including halos with significant angular 
momentum and massive halos with shallow potentials,
and (5) environmental effects such as background potentials.

We will present a more detailed description of 
these simulations (Barton {\it et al.} 1998a) as well as the
data (Barton {\it et al.} 1998b). 
We plan to enlarge our simulation study to explore a
greater portion of parameter space; we will study the frequency 
and duration of interaction-induced kinematic features, as well 
as their dependence on orbital inclination and energy (including 
collisions that do not result in mergers) and starting model parameters. 

Our principle result is that we reproduce some of
the interesting features seen in the rotation curves of galaxies
in our sample.  
These features in our simulated rotation curves
are associated with tidal shearing and subsequent infall.  
Even without gas in our simulations, we 
produce two-component kinematic structures similar to dips
in observed rotation curves
associated with both bars and circumnuclear gas disks 
(Rubin {\it et al.} 1997).  

During interactions, rotation curves derived  from line-of-sight 
velocity measurements deviate from circular speed curves.  
Thus, our simulations illustrate the danger of
inferring galaxy potentials from observed kinematics when interactions
may be a factor.

\begin{acknowledgements}

We thank Antonaldo Diaferio, Mike Kurtz and Lars Hernquist for very
useful discussions and aid.  EJB received support from a Harvard Merit
Fellowship.  BCB acknowledges funding from NSF Grant PHY 95-07695. We
are grateful to the Caltech Center for Advanced Computing Research and
NASA Offices of Space Sciences, Aeronautics, and Mission to
Planet Earth for providing computing resources.

\end{acknowledgements}

\begin{references}

\reference {Arp66} Arp, H. 1966, \apjs, 14, 1
\reference {BarHer92a} Barnes, J. E., \& Hernquist, L. 1992, \araa, 30, 705
\reference {BarHut86} Barnes, J. E., \& Hut, P. 1986, Nature, 324, 446 
\reference {Baretal98b} Barton, E. J., {\it et al.} 1998a, in preparation
\reference {Baretal98c} Barton, E. J., {\it et al.} 1998b, in preparation
\reference {B98a} Borne, K. D. 1988a, \apj, 330, 38
\reference {B98a} Borne, K. D. 1988b, \apj, 330, 61
\reference {B76} Byrd, G. G. 1976, \apj, 208, 688
\reference {BK90} Byrd, G. G., \& Klaric, M. 1990, \aj, 99, 1461
\reference {CheSalTer93} Chengalur, J. N., Salpeter, E. E., \& Terzian, Y. 1993, \apj, 419, 30
\reference {CheSalTer94} Chengalur, J. N., Salpeter, E. E., \& Terzian, Y. 1994, \aj, 107, 1984
\reference {CheSalTer96} Chengalur, J. N., Salpeter, E. E., \& Terzian, Y. 1996, \apj, 461, 546
\reference {Com95} Combes, F., Rampazzo, R., Bonfanti, P. P., Pringniel, P., \& Sulentic, J. W. 1995, \aap, 297, 37
\reference {DubMihHer96} Dubinski, J., Mihos, J. C., \& Hernquist, L. 1996, \apj, 462, 576
\reference {DubHerMih97} Dubinski, J., Hernquist, L., \& Mihos, J. C. 1997, preprint
\reference {E95} Elmegreen, D. M., Kaufman, M., Brinks, E., Elmegreen, B. G., \& Sundin, M. 1995a, \apj, 453, 100
\reference {E95} Elmegreen, B. G., Sundin, M., Kaufman, M., Brinks, E., \& Elmegreen, D. M. 1995b, \apj, 453, 139
\reference {Fal98} Falco, E. E., Kurtz, M. J., Geller, M. J., Huchra, J. P., Peters, J., Berlind, P., Tokarz, S., \& Elwell, B. 1998, in preparation
\reference {GelHuc89} Geller, M. J., \& Huchra, J. P. 1989, Science, 246, 897
\reference {Her88} Hernquist, L. 1987, ApJS, 64, 715
\reference {HibMih95} Hibbard, J. E., \& Mihos, J. C. 1995, \aj, 110, 140
\reference {Huc98} Huchra, J. P., Vogeley, M. S., \& Geller, M. J. 1998, submitted
\reference {Kar72} Karachentsev, I. D. 1972, Soobsch. Spets. Astrof. Obs., 7, 3
\reference {Kar87} Karachentsev, I. D. 1987, Dvoinye Galaktiki (Nauka: Moscow)
\reference {Kee93} Keel, W. C. 1993, \aj, 106, 1771
\reference {Kee96} Keel, W. C. 1996, \apjs, 106, 27
\reference {KuiDub95} Kuijken, K., \& Dubinski, J. 1995, \mnras, 277, 1341
\reference {MarMol96} M\'{a}rquez, I., \& Moles, M. 1996, \aaps, 120
\reference {MihHer96} Mihos, J. C., \& Hernquist, L. 1996, \apj, 464, 641
\reference {NegWhi83} Negroponte, J., \& White, S. D. M. 1983, \mnras, 205, 1009
\reference {Nob87} Noguchi, M. 1991, \mnras, 251, 360
\reference {RubKenYou97} Rubin, V. C., Kenney, J. D. P., \& Young, J. S. 1997, \aj, 113, 1250
\reference {SanMir96} Sanders, D. B., \& Mirabel, I. F. 1996, \araa, 34, 749
\reference {SprWhi98} Springel, V., \& White, S. D. M. 1998, \mnras, submitted
\reference {TooToo72} Toomre, A., \& Toomre, J. 1972, \apj, 178, 623
\reference {Tur76a} Turner, E. L. 1976a, \apj, 208, 20
\reference {Tur76b} Turner, E. L. 1976b, \apj, 208, 304
\reference {Vor59} Vorontsov-Vel'yaminov, B. A. 1959, Atlas and Catalog of Interacting
Galaxies (Moscow: Sternberg State Astronomical Institute)

\end{references}
\end{document}